\documentstyle[preprint,aps,eqsecnum]{revtex}

\newcommand{\lbar}{\overline}
\newcommand{\spa}{\vspace{.25cm}}
\newcommand{\beq}{\begin{equation}}
\newcommand{\eeq}{\end{equation}}
\newcommand{\beqs}{\begin{eqnarray}}
\newcommand{\eeqs}{\end{eqnarray}}
\newcommand{\sg}{{\bf 1}}
\newcommand{\db}{{\bf 2}}
\newcommand{\tb}{{\bf 3}}
\newcommand{\dtd}{$\db \otimes \db$}
\newcommand{\ttt}{$\tb \otimes \tb$}
\newcommand{\ta}{\bar \tb}
\newcommand{\tatta}{$\ta \otimes \ta$}
\newcommand{\SUSYwoR}{SUSY $\not\!\!R_p$}
\newcommand{\calS}{{\cal S}}
\newcommand{\calV}{{\cal V}}

\newcommand{\GeV}{\mbox{\,{\rm GeV}}}

\newcommand{\bL}{\mbox{\boldmath $L$}}
\newcommand{\sbL}{\mbox{\boldmath \scriptsize$L$}}
\newcommand{\DDbar}{$D^0$--$\bar D^0$}
\newcommand{\KKbar}{$K^0$--$\bar K^0$}

\def\lsim{\ \rlap{\raise 3pt \hbox{$<$}}{\lower 3pt \hbox{$\sim$}}\ }
\def\gsim{\ \rlap{\raise 3pt \hbox{$>$}}{\lower 3pt \hbox{$\sim$}}\ }

\def\hepph#1{{\tt hep-ph/#1}}
\def\hepex#1{{\tt hep-ex/#1}}

\begin{document}
\draft
{\tighten
\preprint{\vbox{\hbox{WIS-99/32/Sept-DPP}
                \hbox{hep-ph/9909391}
                \hbox{September 1999}}}

\title{~\\~\\ New Physics Effects in Doubly Cabibbo Suppressed $D$ Decays}
\author{Sven Bergmann and Yosef Nir}
\address{ \vbox{\vskip 0.truecm}
  Department of Particle Physics \\
  Weizmann Institute of Science, Rehovot 76100, Israel}

\maketitle

\begin{abstract}

\noindent
The most sensitive experimental searches for \DDbar\ mixing use $D^0
\to K^+ \pi^-$ decays. It is often assumed that effects of New Physics
and, in particular, CP violation, can appear through the mixing, while
the $c \to d u \bar s$ decay amplitude cannot have significant
contributions from New Physics and is, therefore, CP conserving to a
good approximation. We examine this assumption in two ways. First, we
calculate the contributions to the decay in various relevant models of
New Physics: Supersymmetry without $R$-parity, multi-scalar models,
left-right symmetric models, and models with extra quarks. We find
that phenomenological constraints imply that the New Physics
contributions are indeed small compared to the standard model doubly
Cabibbo suppressed amplitude. Second, we show that many of our
constraints hold model-independently. We find, however, one case where
the model-independent bound is rather weak and a CP violating
contribution of order 30\% is not excluded.

\end{abstract}
} 

\newpage

\section{Introduction}

The decay $\bar D^0 \to K^+ \pi^-$ proceeds via the quark sub-process
$\bar c \to d \bar u \bar s$ and is Cabibbo favored:
\beq \label{CaFa}
A^{\rm SM}(\bar D^0 \to K^+ \pi^-) \propto G_F |V_{cs}V_{ud}|.
\eeq
The decay $D^0 \to K^+ \pi^-$ proceeds via the quark sub-process
$c \to d u \bar s$ and is doubly Cabibbo suppressed:
\beq \label{CaSu}
A^{\rm SM}(D^0 \to K^+ \pi^-) \propto G_F |V_{cd}V_{us}|.
\eeq
If \DDbar\ mixing is large, then there could be a significant
contribution to the latter from $D^0 \to \bar D^0 \to K^+ \pi^-$,
where the second stage is Cabibbo favored. The most sensitive
experimental searches for \DDbar\ mixing use indeed this process. The
fact that the first-mix-then-decay amplitude gives a different time
dependence than the direct decay allows experimenters to distinguish
between the two contributions and to set unambiguous upper bounds on
the mixing.

The standard model (SM) prediction for \DDbar\ mixing,
$(\Delta m_D/m_D)_{\rm SM} \sim 10^{-16}$
\cite{chen,PST,DaKu,Wolf,DGHT,ChSh,CNP,Geor,ORS,Pakv,Kaed,Petr,ABNP,GoPe},
is well below the present experimental sensitivity,
$(\Delta m_D/m_D)_{\rm exp} < 8.5 \times 10^{-14}$
\cite{aitaA,aitaB,alepD,aitaC,cleoD}. If mixing is discovered within
an order of magnitude of present bounds, its theoretical explanation
will require contributions from New Physics.  Even more convincing
evidence for New Physics will arise if CP violation plays a role in
the $D^0 \to K^+ \pi^-$ decay~\cite{BSN,Wolint}. The reason is that,
while the calculation of the total rate suffers from large hadronic
uncertainties related to the long distance contributions, the SM
prediction that there is no CP violation is very safe since it is only
related to the fact that the third generation plays almost no role in
both the mixing and the decay.

Most if not all present analyses of the search for \DDbar\ mixing
through $D\to K\pi$ decays make the assumption that the New Physics
can affect significantly the mixing but not the decay. This is a
plausible assumption. The SM contribution to the mixing is highly
suppressed because it is second order in $\alpha_W$ and has a very
strong GIM suppression factor, $m_s^4/(M_W m_c)^2$. The mixing is then
sensitive to New Physics which could contribute at tree level (as in
multi-scalar models), or through strong interactions (as in various
supersymmetric models), etc.  On the other hand, the SM contribution
to the decay is through the tree-level $W$-mediated diagram.  One does
not expect that New Physics could give competing contributions.

Yet, since the decay in question is doubly Cabibbo suppressed, one may
wonder if indeed the assumption that it gets no New Physics
contributions is safe.  It is the purpose of this work to test this
assumption in a more concrete way. (For previous work on related
processes, see~\cite{YOR,BiYa,LiXi}.) We examine various reasonable
extensions of the standard model with new tree level contributions to
the decay. For each model, we present the relevant phenomenological
constraints and find an upper bound on the new contributions to $D^0
\to K^+ \pi^-$.

From (\ref{CaFa}) and (\ref{CaSu}) we get the following (naive)
estimate for the ratio of amplitudes:
\beq \label{Ratio}
\left| A^{\rm SM}(D^0 \to K^+ \pi^-) \over
       A^{\rm SM}(\bar D^0 \to K^+ \pi^-) \right| \sim
\left |V_{cd} V_{us} \over V_{cs} V_{ud} \right| \sim 0.05 \,.
\eeq
The value of this ratio from the recent CLEO results~\cite{cleoD} is
about $0.058$. Thus, if New Physics contributions to $D^0 \to K^+
\pi^-$ are to compete with the doubly Cabibbo suppressed SM amplitude,
the corresponding effective New Physics coupling $G_N$ should satisfy
\beq
G_N \gsim 10^{-2}\ G_F \,.
\eeq
In Section~\ref{specificmodels} we investigate if this is possible in
various New Physics scenarios. In Section~\ref{modelindependent} we
study model-independent bounds on new tree-level contributions to $D^0
\to K^+\pi^-$.  We conclude in Section~\ref{conclusions}.

\section{Specific Models}
\label{specificmodels}

\subsection{Supersymmetry without $R$-parity}
\label{SUSYwoR}
Supersymmetry without $R$-parity ($R_p$) predicts new tree diagrams
contributing to the decay. The lepton number violating terms
$\lambda^\prime_{ijk} L_i Q_j d^c_k$ give a slepton-mediated
contribution with an effective coupling:
\beq \label{CaSuRp1}
G_N^{\lambda'} = {\lambda^{\prime}_{21k}\lambda^{\prime*}_{12k}
 \over {4 \sqrt2 M^2(\tilde\ell_{Lk}^-)}} \,.
\eeq
These couplings are severely constrained by \KKbar\ mixing
(see {\it e.g.} \cite{Bhat}):
\beq \label{Boulp}
{\lambda^{\prime}_{21k}\lambda^{\prime*}_{12k}} \lsim 10^{-9}
 ~~~({\rm for}\ M(\tilde\ell_{Lk}^-) = 100\ \GeV).
\eeq
This rules out any significant contribution to $D^0 \to K^+ \pi^-$
from slepton exchange in models of $R_p$ violation:
\beq \label{Glambdaprime}
{G_N^{\lambda'} \over G_F|V_{cd}V_{us}|} \lsim 3 \times 10^{-8} \,.
\eeq
The baryon number violating terms $\lambda''_{ijk} u^c_i d^c_j d^c_k$
give a squark-mediated contribution with an effective coupling
\beq \label{CaSuRp2}
G_N^{\lambda''} = {\lambda''_{113} \lambda''^*_{223}
 \over {4 \sqrt2 M^2(\tilde b_R)}} \,.
\eeq
The $\lambda''_{223}$ couplings is only constrained by requiring that
it remains in the perturbative domain up to the unification scale and
could be of order unity~\cite{BrRo}.  The $\lambda''_{113}$ coupling
is, however, severely constrained by the upper bound on $n-\bar n$
oscillations~\cite{GoSh}:
\beq \label{Boulpp}
|\lambda''_{113}| \lsim 10^{-4}
 ~~~({\rm for}\ M(\tilde q) = 100\ \GeV).
\eeq
This rules out a significant contribution to $D^0 \to K^+ \pi^-$ from
squark exchange in models of $R_p$ violation:
\beq
{G_N^{\lambda''}\over G_F|V_{cd}V_{us}|} \lsim 3 \times10^{-3} \,.
\eeq


\subsection{Multi-Scalar Models}
\label{2HDM}

Extensions of the scalar sector, beyond the single Higgs doublet of
the SM, predict new tree diagrams contributing to the decay.

In two Higgs doublet models (2HDM) with natural flavor conservation,
there is a charged Higgs ($H^\pm$) mediated contribution. The
trilinear coupling of the physical charged Higgs to the $u_i \, \bar
d_j$ bilinear is
\beq \label{couplingHDM}
-{\cal L}_{H^\pm} = {i g \over \sqrt{2} m_W}\, \lbar{u_{i}} \,
\left[m_{u_i} \cot \beta \, P_L +
      m_{d_j} \tan \beta \, P_R \right] V_{ij} \, d_{j} H^{+} + h.c. \,,
\eeq
where $m_W$ is the mass of the $W$-boson, $m_q$ is the mass of the
quark $q$, $\tan \beta = v_u / v_d$ is the ratio of vevs and $P_{R,L}
= (1 \pm \gamma_5)/2$.  It follows that the charged Higgs mediated
contribution is also doubly Cabibbo suppressed. Then, for large $\tan
\beta$, the suppression with respect to the SM contribution is given
by
\beq \label{Htwolargeb}
{G_N^{H^\pm} \over G_F|V_{cd}V_{us}|} \simeq
{m_d \, m_s \tan^2 \beta \over M_{H^\pm}^2} \lsim 4 \times 10^{-4} \,.
\eeq
To obtain the upper bound, we used the constraint from $b \to c \,
\tau \, \nu$~\cite{GrLi,GHN}:
\beq \label{tanbeta}
\tan\beta \lsim 0.5 \left( {M_{H^\pm} \over \GeV} \right) \,,
\eeq
and the ranges of quark masses given in Ref.~\cite{PDG}.  For $\tan
\beta \simeq 1$ we have
\beq \label{Htwosmallb}
{G_N^{H^\pm} \over G_F|V_{cd}V_{us}|} \simeq {m_s \, m_c \over M_{H^\pm}^2}
 \lsim 10^{-4} \,.
\eeq
To obtain the upper bound, we used $M_{H^\pm} \gsim
54.5$~GeV~\cite{PDG}.  Thus there are no significant contributions to
$D^0 \to K^+ \pi^-$ from charged Higgs exchange within 2HDM.

Multi Higgs doublet models with natural flavor conservation but with
more than two Higgs doublets have parameters that are less constrained
and, in particular, provide new sources of CP violation. There are
several charged scalars that can mediate the $D^0 \to K^+ \pi^-$
decay. If we take the simplest case that only one of them contributes
in a significant way (see {\it e.g.} \cite{Gros}), then its couplings
are similar to those of Eq.~(\ref{couplingHDM}) except that
$\tan\beta$ and $\cot\beta$ are replaced by, respectively, $X$ and
$Y$. In general, $X$ and $Y$ are complex and, moreover, $|XY| \neq 1$.
Eq.~(\ref{Htwolargeb}) is modified:
\beq
{G_N^{H^\pm} \over G_F|V_{cd}V_{us}|} \simeq
{m_d \, m_s |X|^2 \over M_{H^\pm}^2} \lsim 10^{-2} \,.
\eeq
To obtain the upper bound, we used the perturbativity bound $|X| \lsim
130$~\cite{BHP,Gros} and the lower bound on $M_{H^\pm}$.  Note that
this contribution is not only constrained to be small, but also it
carries no new CP violating phase. In contrast, the new contribution
that replaces that of Eq.~(\ref{Htwosmallb}) does carry a new phase:
\beq
{G_N^{H^\pm} \over G_F|V_{cd}V_{us}|} \simeq
 {m_s \, m_c YX^*\over M_{H^\pm}^2}
\lsim 3 \times 10^{-4} \,.
\eeq
To obtain the upper bound, we used the constraint from $b \to s
\gamma$, $|XY| \lsim 4$~\cite{Gros}. The bound on a CP violating
contribution is even somewhat stronger, since the measurement of $b
\to s \gamma$ gives ${\cal I}m(X^*Y) \lsim 2$~\cite{GrNi}. In any
case, the contribution from charged Higgs exchange in multi Higgs
doublet models is, at most, at the percent level. The CP violating
part of this contribution is at most of order $10^{-4}$\,.

It is possible that Yukawa couplings are naturally suppressed by
flavor symmetries rather than by natural flavor conservation
\cite{ChSh}.  In such a framework, there is a contribution to $D^0 \to
K^+ \pi^-$ from neutral scalar exchange. To estimate these
contributions, we use the explicit models of Ref. \cite{LNS}. Here, a
horizontal $U(1)_{\cal H}$ symmetry is imposed. At low energies, the
symmetry is broken by a small parameter $\lambda$ (usually taken to be
of the order of the Cabibbo angle, $\lambda \sim 0.2$), leading to
selection rules. The scalar sector consists of two Higgs doublets,
$\phi_u$ and $\phi_d$, and a single scalar singlet $S$.  The effective
coupling of the $S$ scalar to quarks is given by
\beq \label{Yukawa}
-{\cal L}_S = Z^{q}_{ij}\,S\,\lbar{q_{iR}} \, q_{jL} + h.c.
 ~~~(q=u,d,\ \ i,j=1,2,3).
\eeq
The order of magnitude of $Z^{q}_{ij}$ is determined by the selection
rules related to the broken flavor symmetry:
\beq \label{YukawaAHS}
Z^q_{ij} \sim {M^q_{ij}\over\langle S\rangle} \,, \ \ \
M^q_{ij} \sim \lambda^{{\cal H}(q_{jL})+{\cal H}(\lbar{q_{iR}})
+{\cal H}(\phi_q)}\langle \phi_q\rangle \,.
\eeq
The horizontal charges ${\cal H}$ of the quark and Higgs fields are
determined by the physical flavor parameters:
\beqs \label{Hrel}
|V_{ij}| &\sim& \lambda^{{\cal H}(q_{iL}) - {\cal H}(q_{jL})}, \\
m(q_i) &\sim& \lambda^{{\cal H}(q_{iL})
 + {\cal H}(\lbar{q_{iR}})+{\cal H}(\phi_q) }\bigl<\phi_q\bigr>.
\nonumber
\eeqs
Using~(\ref{Hrel}) we can express the suppression of the relevant
Yukawa couplings in terms of the quark masses and mixing angles:
\beq \label{YukawaSup}
|Z^u_{uc}| \sim {m_c |V_{12}|\over \langle S\rangle} \, ,~~~
|Z^u_{cu}| \sim {m_u \over \langle S\rangle|V_{12}|} \, ,~~~
|Z^d_{ds}| \sim {m_s |V_{12}|\tan\beta\over \langle S\rangle} \, ,~~~
|Z^d_{sd}| \sim {m_d \tan\beta\over \langle S\rangle|V_{12}|} \, .
\eeq
These couplings give rise to various operators that induce $c \to u d
\lbar{s}$ at tree level.  For the leading contributions, we find
\beq \label{Hops}
{G_N^S \over G_F|V_{cd}V_{us}|} \sim
{m_c m_s \tan\beta \over \langle S \rangle^2} {m_W^2\over m_S^2}
\lsim 5\times10^{-3} \,.
\eeq
To obtain the upper bound, we used $\tan\beta\lsim130$ and the very
conservative bound $\langle S\rangle \sim m_S > m_W$. Other
models~\cite{ChSh,AHR,HaWe} give a similar or even stronger
suppression.  We conclude that there are no significant contributions
to $D^0 \to K^+ \pi^-$ from neutral Higgs exchange within multi-scalar
models with approximate flavor symmetries.


\subsection{Left-Right Symmetric Models}
\label{LRSM}

Left-right symmetric (LRS) models predict new tree-level
contributions, mediated by the $W_R$ gauge bosons. The relevant
interactions are given by
\beq \label{LRint}
-{\cal L}_{CC} =
{g_R \over \sqrt{2}} \, \lbar{u_{iR}} \, \gamma_\mu V^R_{ij} \, d_{jR}
 W^{\mu+}_R + h.c. \,,
\eeq
where $V^R$ is the mixing matrix for the right-handed quarks. For a
general model of an extended electroweak gauge group $G=SU(2)_L\times
SU(2)_R \times U(1)_{B-L}$, the interactions of Eq.~(\ref{LRint}) lead
to
\beq \label{RatioLR2}
{G_N^{W_R} \over G_F \left| V_{cd} V^*_{us} \right|} =
 {g_R^2 \over g_L^2}{m_{W_L}^2 \over m_{W_R}^2}
 \left| V^R_{cd} V^{R*}_{us} \over V_{cd} V^*_{us} \right|\,.
\eeq
However, in left-right symmetric models, an extra discrete symmetry is
imposed. It leads to the relation $g_L=g_R$ and, in models of
spontaneous CP violation or of manifest left-right symmetry, to
$|V_{ij}| = |V_{ij}^R|$. Then Eq.~(\ref{RatioLR2}) is simplified:
\beq \label{RatioLR}
{G_N^{\rm LRS} \over G_F \left| V_{cd} V^*_{us} \right|} =
 {m_{W_L}^2 \over m_{W_R}^2} \lsim {1 \over 430} \,,
\eeq
where the upper bound comes from the $\Delta m_K$
constraint~\cite{BBS}.

In $SU(2)_L \times SU(2)_R \times U(1)_{B-L}$ models where $V$ and
$V_R$ are independent mixing matrices, it is possible to avoid the
$\Delta M_K$ constraints~\cite{LaSa,LoWy}. This is done by fine tuning
the relevant entries in $V_R$ to be very small. In particular, it was
shown that in such a framework there could be interesting implications
on CP violation in the $B$ system~\cite{LoWy}.  However, as concerns
the $D^0 \to K^+ \pi^-$ decay, the situation is different: the {\it
  same} mixing elements that contribute to $D^0 \to K^+ \pi^-$, that
is $V^R_{cd}V^{R*}_{us}$, contribute also to $K-\bar K$ mixing.  If
they are switched off, to avoid the $\Delta m_K$ constraint, the new
contribution to $D^0 \to K^+ \pi^-$ vanishes as well.  One can see
that independently of the details of the model by noticing that the
$G_N^{W_R}$ effective coupling of Eq.~(\ref{RatioLR2}) can be combined
with the flavor-changing $G_F V_{cd} V^*_{us}$ coupling of the SM to
produce a contribution to $K-\bar K$ mixing. Indeed, one finds for the
CP conserving contribution~\cite{LaSa}:
\beq \label{ReWRft}
{\cal R}e \left({G_N^{W_R} \over G_F V_{cd} V^*_{us}} \right) 
\lsim 0.2 \,,
\eeq
and for the CP violating contribution~\cite{LoWy}:
\beq \label{ImWRft}
{\cal I}m \left({G_N^{W_R} \over G_F V_{cd} V^*_{us}} \right) 
\lsim 0.002 \,.
\eeq
We learn that in such fine-tuned models, the $W_R$-mediated
contribution to the decay rate could be non-negligible, but the CP
violating contribution is very small.


\subsection{Extra Quarks in SM Vector-Like Representations}
\label{extraquarks}

In models with non-sequential (`exotic') quarks, the $Z$-boson has
flavor changing couplings, leading to a $Z$-mediated contribution to
the $D^0 \to K^+ \pi^-$ decay. For example, in models with additional
up quarks in the vector-like representation $(\tb,\sg,+2/3) \oplus
(\ta,\sg,-2/3)$ and additional down quarks in the vector-like
representation $(\tb,\sg,-1/3) \oplus (\ta,\sg,+1/3)$, the flavor
changing $Z$ couplings have the form
\beq \label{vectorcoupling}
-{\cal L}_Z=
{g\over2\cos\theta_W}\left(U^u_{ij} \, \lbar u_{Li} \gamma_\mu u_{Lj} -
U^d_{ij} \, \lbar d_{Li} \gamma_\mu d_{Lj} \right) Z^\mu + h.c. \,.
\eeq
Here, $U^q = V_L^{q\dagger}\,{\rm diag}(1,1,1,0)\,V_L^q$, where
$V_L^q$ is the $4 \times 4$ diagonalizing matrix for $M_q M_q^\dagger$
\linebreak ($M_q$ being the quark mass matrix).  The flavor changing
couplings are constrained by $\Delta M_K$ and $\Delta M_D$:
\beq \label{Uijcon}
|U^d_{sd}| \lsim 2 \times 10^{-4} \,,\ \ \
|U^u_{cu}| \lsim 7 \times 10^{-4} \,.
\eeq
The resulting effective four fermi coupling is given by
\beq \label{CaSuZ}
{G_N^Z\over G_F|V_{cd}V_{us}|} 
\simeq {|U^d_{sd} U^u_{cu}|\over|V_{cd}V_{us}|}
\lsim 3 \times 10^{-6} \,.
\eeq
The same bound applies for the case of vector-like quark doublets,
$(\tb,\db,+1/6) \oplus (\ta,\db,-1/6)$\,. The flavor changing $Z$
couplings are to right-handed quarks, with a mixing matrix $U^q =
V_R^{q\dagger}\,{\rm diag}(0,0,0,1)\,V_R^q$. Here $V_R^q$ is the
$4\times4$ diagonalizing matrix for $M_q^\dagger M_q$.

We learn that a significant contribution to $D^0 \to K^+ \pi^-$ from
$Z$-mediated flavor changing interactions is ruled out.

\section{Model Independent Analysis}
\label{modelindependent}

We have seen that the contributions to $D^0 \to K^+ \pi^-$ in various
reasonable extensions of the SM cannot compete with the $W$-mediated
process. Still, it would be useful if one could show {\it
  model-independently}\/ that CP violation in decay can be neglected.
We try to accomplish this task for all possible tree level
contributions to the $D^0 \to K^+ \pi^-$ decay.  Our analysis proceeds
as follows~\cite{BeGr}: We first list all relevant (anti)quark
bilinears and their transformation properties under the SM gauge group
${\cal G}_{SM} = SU(3)_C \times SU(2)_L \times U(1)_Y$. If the two
quarks have the same (opposite) chirality, they couple to a scalar
(vector) boson. Altogether there are ten possible bilinears (plus
their hermitian conjugates) that are shown in Tab.~1\,.  Here $Q$
denotes the left-handed quark doublet, $q=u,d$ denote the right-handed
quark singlets, and the superscript $c$ refers to the respective
antiquarks. The examples given in the last column refer to the models
discussed in Section~\ref{specificmodels}.

In general, the presence of a heavy boson ${\cal B}$ that couples to
any of the above quark bilinears $B_{ij}$ with trilinear couplings
$\lambda^{\cal B}_{ij}$, where $i, j = 1, 2, 3$ refer to the quark
flavors, gives rise to the four quark operator $B_{ij}^\dagger B_{kl}$
with the effective coupling
\beq \label{GN}
G^{\cal B}_N = C_{CG}
{{\lambda^{\cal B}_{ij}}^* \lambda^{\cal B}_{kl}
\over {4 \sqrt2 M_{\cal B}^2}} \,,
\eeq
at energy scales well below the mass $M_{\cal B}$. ($C_{CG}$ is the
appropriate Clebsch-Gordan coefficient.) For intermediate diquarks, we
only discuss color triplets. The discussion of color sextets follows
similar lines.

\begin{center}
\begin{tabular}{| c || c  | c | c || c || c |}
\hline
~Bilinear $B$~ & ~$SU(3)_C$~ & ~$SU(2)_L$~ & ~$Y$~
               & ~Couples to Boson ${\cal B}$~ & Example  \\
\hline \hline
$Q d^c$    & $\sg$  & $\db$ & ~$1/2$~  & $\calS(\sg,\db,-1/2)$ &
 $\tilde L$   (\SUSYwoR) \cr
\hline
$u^c d^c$  & \tatta & $\sg$ & ~$-1/3$~ & $\calS(\ta,\sg,1/3)$ &
 $~\tilde d^c$ (\SUSYwoR)~ \cr
\hline
$Q u^c$    & $\sg$  & $\db$ & ~$-1/2$~ & $\calS(\sg,\db,1/2)$ &
 $H_u$        (2HDM) \cr
\hline
$Q Q$      & \ttt   & \dtd  & ~$1/3$~  & ~$\calS(\tb,\bL,-1/3)$
                                                 [\bL=\sg,\tb]~ & \cr
\hline
\hline
$u d^c$    & $\sg$  & $\sg$ & ~$1$~    & $\calV(\sg,\sg,-1)$ &
 $W_R$ (LRS) \cr
\hline
$q q^c$    & $\sg$  & $\sg$ & ~$0$~    & $\calV(\sg,\sg,0)$ & \cr
\hline
$Q d$      & \ttt   & $\db$ & ~$-1/6$~ & $\calV(\tb,\db,1/6)$ & \cr
\hline
$Q u$      & \ttt   & $\db$ & ~$5/6$~  & $\calV(\tb,\db,-5/6)$ & \cr
\hline
$Q Q^c$    & $\sg$  & \dtd  & ~$0$~    & $\calV(\sg,\bL,0)$
                                                [\bL=\sg,\tb]~ &
$Z$ (extra $q$'s) \cr
\hline
\end{tabular}

\spa
Tab.~1: Quark-(Anti)Quark Bilinears \\
\end{center}
\spa

In order to predict the rates of the relevant hadronic process one
would need to take into account QCD corrections as well as the
hadronic matrix elements. Since we are mainly interested in ratios
between the rates due to New Physics and those from the SM,
using~(\ref{GN}) is sufficient to obtain an order-of-magnitude
estimate for such ratios.

The first entry in Tab.~1 is realized in supersymmetric models without
$R_p$ (\SUSYwoR): \linebreak $\calS(\sg,\db,-1/2)$ is the slepton
doublet $\tilde{L_k}$, with $\lambda^{\calS(\sg,\db,-1/2)}_{ij} =
\lambda'_{ijk}\,$. As we have pointed out in Section~\ref{SUSYwoR},
non-vanishing $\lambda^{\calS(\sg,\db,-1/2)}_{12}$ and
$\lambda^{\calS(\sg,\db,-1/2)}_{21}$ give rise not only to tree-level
contributions to $D^0 \to K^+ \pi^-$ but also to \KKbar\ mixing, which
severely constraints the effective coupling $G_N$. In this case the
bound arises only from the presence of the trilinear couplings and
supersymmetry does not play a role. The bound in~(\ref{Glambdaprime})
is then model-independent.

The second entry in Tab.~1 is also realized in supersymmetric models
without $R_p$: $\calS(\tb,\sg,1/3)$ is the down squark
$\tilde{d^c_k}$, with $\lambda^{\calS(\tb,\sg,1/3)}_{ij} =
\lambda''_{ijk}$.  For the $\lambda''$ coupling, however, the
constraint comes from the upper bound on $n - \bar n$ oscillations: to
violate baryon number but conserve strangeness or beauty, an internal
loop with charginos is required~\cite{GoSh}.  Supersymmetry does play
a role in the bound on $\lambda''_{113}$, and the bound does not hold
for a generic $\lambda^{\calS(\tb,\sg,1/3)}_{11}$.  More generally,
there is no strong model-independent bound on any diagonal
$\lambda^{\calS(\tb,\sg,1/3)}_{ii}$ coupling. The bound on the scale
of compositeness~\cite{PDG}, $\Lambda(qqqq) \gsim 1.6$~TeV, suggests a
bound for the $i=1$ case, $|\lambda^{\calS(\tb,\sg,1/3)}_{11}| \lsim
0.2$ which implies $G_N^{\calS(\tb,\sg,1/3)} \lsim 0.3~G_F$ (assuming
$|\lambda^{\calS(\tb,\sg,1/3)}_{22}| \sim 1$). We learn then that one
could construct models which incorporate color-triplet weak-singlet
scalars where there is a large CP violating contribution to $D^0 \to
K^+ \pi^-$.

The coupling of $Q u^c$ to $\calS(\sg,\db,1/2)$ appears in the two
Higgs doublet model with natural flavor conservation, as discussed in
Section~\ref{2HDM}. In this model, the effective coupling is
suppressed by the quark masses and the CKM matrix elements. But also
if the doublet $\calS(\sg,\db,1/2)$ is unrelated to the generation of
the quarks masses, one can derive a model-independent bound, which
only relies on the $SU(2)_L$ symmetry: Non-vanishing
$\lambda^{\calS(\sg,\db,1/2)}_{12}$ and
$\lambda^{\calS(\sg,\db,1/2)}_{21}$ give not only a charged scalar
mediated contribution to $D^0 \to K^+ \pi^-$, but also a neutral
scalar mediated contribution to \DDbar\ mixing. We are assuming that
the New Physics takes place at a scale that is comparable to or higher
than the electroweak breaking scale, so that $SU(2)_L$ breaking
effects are not large and the masses of the charged and neutral
scalars are similar~\cite{BGP}. Consequently, the upper bound on
\DDbar\ mixing translates into
\beq
G^{{\calS(\sg,\db,1/2)}_{12}}_N =
{{\lambda^{\calS(\sg,\db,1/2)}_{12}}^* \lambda^{\calS(\sg,\db,1/2)}_{21}
 \over 4 \sqrt2 M_{\calS(\sg,\db,1/2)}^2} \lsim 10^{-7}\ G_F \,,
\eeq
too small to compete with the SM contribution.

The coupling of the $Q Q$ bilinear to a scalar field could induce $D^0
\to K^+ \pi^-$ if the flavor diagonal entries,
$\lambda^{\calS(\tb,\sbL,-1/3)}_{11}$ and
$\lambda^{\calS(\tb,\sbL,-1/3)}_{22}$, are non-zero. For an $SU(2)_L$
singlet ($\bL=\sg$), the coupling is flavor anti-symmetric and
therefore $\lambda^{\calS(\tb,\sg,-1/3)}_{ii}=0$. For an $SU(2)_L$
triplet ($\bL=\tb$), the coupling is flavor symmetric and
$\lambda^{\calS(\tb,\tb,-1/3)}_{ii} \neq 0$ is possible. (For scalar
$SU(3)_C$ sextets the situation would be reversed.) However, while the
$Q_{EM}=-1/3$ component mediates $D^0 \to K^+ \pi^-$, the
$Q_{EM}=+2/3$ component induces \KKbar\ mixing and the $Q_{EM}=-4/3$
component induces \DDbar\ mixing.  We find:
\beq
G^{{\calS(\tb,\tb,-1/3)}_{12}}_N =
{\lambda^{\calS(\tb,\tb,-1/3)}_{11}\lambda^{\calS(\tb,\tb,-1/3)}_{22}
\over 4 \sqrt2 M^2_{\calS(\tb,\tb,-1/3)}} \lsim 10^{-7}\ G_F \,,
\eeq
too small to compete with the SM contribution.

Among the vector bosons listed in Tab.~1 we already encountered
specific examples for the color singlets $\calV(\sg,\sg,-1)$ ($W_R^-$
in LRS models) and $\calV(\sg,\tb,0)$ ($Z$-induced FCNCs due to extra
quarks). The discussion we presented in Section~\ref{LRSM} can be
generalized to any theory that contains a vector boson
$\calV(\sg,\sg,-1)$ that couples to $u d^c$ as in~(\ref{LRint}). Note
that the $W_R$ (being a gauge boson) has flavor diagonal couplings in
the flavor basis and only the charged components induce flavor
transitions between the mass eigenstates, while the neutral component
cannot mediate FCNCs. Still, as we have seen in Section~\ref{LRSM},
the contribution from the left-right box diagram to $\Delta M_K$ and
$\epsilon_K$ imposes severe constraints on the $D^0 \to K^+ \pi^-$
amplitude due to $\calV(\sg,\sg,-1)$ exchange and rules out
significant CP violation in the decay.

For the generic coupling $\lambda^{\calV(\sg,\tb,0)}$ we can adopt the
specific result obtained in Section~\ref{extraquarks}.  Since the
argument based on the bounds from \KKbar\ and \DDbar\ oscillations
only used the trilinear couplings (\ref{vectorcoupling}), it can be
generalized to the generic couplings $\lambda^{\calV(\sg,\sg,0)}$.
Because all quark-antiquark bilinears $B_{ij}$ that couple to
$\calV(\sg,\sg,0)$ are gauge-invariant one can easily see that
$\calV(\sg,\sg,0)$ exchange induces not only the flavor-conserving
effective operator $B_{ij}^\dagger B_{ij}$ but also the
flavor-violating operator $B_{ij} B_{ij}$ that gives rise to neutral
meson mixing.

For the remaining vector couplings in Tab.~1, the decay $D^0 \to K^+
\pi^-$ can be induced, if the flavor diagonal couplings
$\lambda^{\calV(\tb,\db,Y)}_{ii}$ for both $i=1$ and 2 are non-zero.
Note that the intermediate vector boson carries color. Since the
respective quark bilinears contain one $SU(2)_L$ doublet the effective
operator that gives rise to $D^0 \to K^+ \pi^-$ is related by an
$SU(2)_L$ rotation to an operator that induces \KKbar\ [for
$\calV(\tb,\db,1/6)$] and \DDbar\ [for $\calV(\tb,\db,-5/6)$]
oscillations at tree level. Since $SU(2)_L$ breaking effects are
small~\cite{BGP}, the data from neutral meson mixing imply $
G_N^{\calV(\tb,\db,1/6)} \lsim 10^{-7}\ G_F \,,$ and
$G_N^{\calV(\tb,\db,-5/6)} \lsim 10^{-6}\ G_F \,,$ ruling out any
significant contribution to $D^0 \to K^+ \pi^-$.

\section{Conclusions}
\label{conclusions}

We have examined well-motivated extensions of the standard model that
give new, tree-level contributions to the $D^0 \to K^+ \pi^-$ decay.
We showed that in all the models that we considered, strong
phenomenological constraints imply that these contributions can be
safely neglected.

We have extended our discussion to a model-independent analysis of all
possible tree level contributions to the decay. We found that there is
only one case where a large contribution to $D^0 \to K^+ \pi^-$ is
possible.  This is the case where two right-handed quarks, $u^cd^c$,
couple to an $SU(2)_L$-singlet scalar, $\calS(\ta,\sg,1/3)$. Such a
coupling is present in SUSY without $R_p$ but in this model the
relevant coupling is constrained by $n - \bar n$ oscillations, ruling
out a contribution that is comparable to the SM doubly Cabibbo
suppressed diagram.

In our analysis, we have implicitly assumed that there are no
significant accidental cancellations between various contributions to
the processes from which we derive our constraints. It is possible to
construct fine-tuned models where there is a large new contribution to
$D^0 \to K^+ \pi^-$ while the related contributions to flavor changing
neutral current processes are small.

We conclude that, in general, the assumption that New Physics effects
could affect the $D^0 \to K^+ \pi^-$ decay and, in particular, its CP
violating part, only through \DDbar\ mixing is a good assumption and
it holds to better than one percent in all the reasonable and
well-motivated extensions of the standard model that we have examined.
One could construct, however, viable (even if unmotivated) models
where there is a new, large [${\cal O}(0.3)$] and possibly CP
violating contribution to the decay.

\acknowledgements

We thank Zvi Lipkin for asking us the questions that led to this
study. Y.\,N. is supported in part by the United States -- Israel
Binational Science Foundation (BSF), by the Israel Science Foundation
founded by the Israel Academy of Sciences and Humanities, and by the
Minerva Foundation (Munich).


{}


\end{document}